\documentclass[times, twocolumn]{aastex63}
\usepackage{CJK}
\usepackage{graphicx}
\usepackage{enumerate}
\usepackage{amssymb, amsmath}
\usepackage{natbib}
\usepackage{color}
\usepackage{ulem}
\usepackage{dblfnote}
\usepackage{appendix}
\usepackage{hyperref}
\usepackage[stable]{footmisc}
\usepackage{comment}
\usepackage{footnote}

\usepackage[nodayofweek]{datetime}
\usepackage{lineno}
\newdateformat{monthyearday}{%
  \THEYEAR\ \monthname[\THEMONTH] \twodigit{\THEDAY}}

\newcommand{\mj}{\ensuremath{\,M_{\rm J}}}

\newcommand{\mum}{$\mu$m}

\begin{document}
\submitjournal{AAS Journals}
\shorttitle{$\copyright$ 2022. All rights reserved.} 
\shortauthors{$\copyright$ 2022. All rights reserved.}

\begin{CJK*}{UTF8}{gbsn}

\title{Utilizing a global network of telescopes to update the ephemeris for the highly eccentric planet HD 80606 b and to ensure the efficient scheduling of JWST}


\correspondingauthor{Kyle A. Pearson, kyle.a.pearson@jpl.nasa.gov}

\author[0000-0002-5785-9073]{Kyle A. Pearson}
\affiliation{Jet Propulsion Laboratory, California Institute of Technology, Pasadena, CA 91125 USA}
\affiliation{Exoplanet Watch, \url{https://exoplanets.nasa.gov/exoplanet-watch/}}

\author[0000-0002-5627-5471]{Charles Beichman}
\affiliation{NASA Exoplanet Science Institute, IPAC, California Institute of Technology, Pasadena, CA 91125 USA}
\affiliation{Jet Propulsion Laboratory, California Institute of Technology, Pasadena, CA 91125 USA}

\author[0000-0003-3504-5316]{B.J. Fulton}
\affiliation{NASA Exoplanet Science Institute, IPAC, California Institute of Technology, Pasadena, CA 91125 USA}

\author[0000-0002-0792-3719]{Thomas M. Esposito}
\affiliation{SETI Institute, Carl Sagan Center, 339 Bernardo Ave, Ste 200, Mountain View, CA 94043 USA}
\affiliation{Unistellar SAS, 19 Rue Vacon, 13001 Marseille, France}
\affiliation{Department of Astronomy, University of California Berkeley, Berkeley, CA 94720 USA}

\author[0000-0001-7547-0398]{Robert T. Zellem}
\affiliation{Jet Propulsion Laboratory, California Institute of Technology, Pasadena, CA 91125 USA}
\affiliation{Exoplanet Watch, \url{https://exoplanets.nasa.gov/exoplanet-watch/}}

\author[0000-0002-5741-3047]{David R. Ciardi}
\affiliation{NASA Exoplanet Science Institute, IPAC, California Institute of Technology, Pasadena, CA 91125 USA}

\author{Jonah Rolfness}
\affiliation{California Institute of Technology, Pasadena, CA 91125 USA}
\affiliation{Jet Propulsion Laboratory, California Institute of Technology, Pasadena, CA 91125 USA}
\affiliation{Exoplanet Watch, \url{https://exoplanets.nasa.gov/exoplanet-watch/}}

\author[0000-0002-5977-5607]{John Engelke}
\affiliation{Jet Propulsion Laboratory, California Institute of Technology, Pasadena, CA 91125 USA}
\affiliation{Raytheon Intelligence, Information, and Services, 300 N Lake Ave, Suite 1120, Pasadena, CA 91101, USA}
\affiliation{Exoplanet Watch, \url{https://exoplanets.nasa.gov/exoplanet-watch/}}

\author[0000-0002-0665-5759]{Tamim Fatahi}
\affiliation{Department of Computer Science, California Polytechnic University, San Luis Obispo USA}
\affiliation{Jet Propulsion Laboratory, California Institute of Technology, Pasadena, CA 91125 USA}
\affiliation{Exoplanet Watch, \url{https://exoplanets.nasa.gov/exoplanet-watch/}}

\author{Rachel Zimmerman-Brachman}
\affiliation{Jet Propulsion Laboratory, California Institute of Technology, Pasadena, CA 91125 USA}
\affiliation{Exoplanet Watch, \url{https://exoplanets.nasa.gov/exoplanet-watch/}}

\author[0000-0001-7801-7425]{Arin Avsar}
\affiliation{Unistellar SAS, 19 Rue Vacon, 13001 Marseille, France}
\affiliation{Department of Astronomy, University of California Berkeley, Berkeley, CA 94720 USA}

\author[0000-0002-6112-7609]{Varun Bhalerao} 
\affiliation{Department of Physics, Indian Institute of Technology Bombay, Powai, 400076, India}

\author{Pat Boyce}
\affiliation{Boyce Research Initiatives and Education Foundation}
\affiliation{Exoplanet Watch, \url{https://exoplanets.nasa.gov/exoplanet-watch/}}

\author{Marc Bretton}
\affiliation{Observatoire des Baronnies Proven{\c c}ales, Route de Nyons, F-05150 Moydans, France}

\author[0000-0001-5248-1705]{Alexandra D. Burnett}
\affiliation{Unistellar Network Citizen Scientist, \url{https://unistellaroptics.com/citizen-science/}}
\affiliation{School of Natural Resources and the Environment, University of Arizona, Tucson, AZ 85721 USA}

\author[0000-0002-0040-6815]{Jennifer Burt}
\affiliation{Jet Propulsion Laboratory, California Institute of Technology, Pasadena, CA 91125 USA}

\author{Martin Fowler}
\affiliation{ExoClock Project, \url{https://www.exoclock.space/}}
\affiliation{Exoplanet Watch, \url{https://exoplanets.nasa.gov/exoplanet-watch/}}

\author{Daniel Gallego}
\affiliation{Exoplanet Watch, \url{https://exoplanets.nasa.gov/exoplanet-watch/}}

\author{Edward Gomez}
\affiliation{Las Cumbres Observatory, 6740 Cortona Drive, Suite 102, Goleta, CA  93117 USA}

\author[0000-0003-4091-0247]{Bruno Guillet}
\affiliation{Unistellar Network Citizen Scientist, \url{https://unistellaroptics.com/citizen-science/}}

\author{Jerry Hilburn} 
\affiliation{Exoplanet Watch, \url{https://exoplanets.nasa.gov/exoplanet-watch/}}

\author{Yves Jongen}
\affiliation{Observatoire de Vaison-La-Romaine, D{\'e}partementale 51, pr{\`e}s du Centre Equestre au Palis—F-84110 Vaison-La-Romaine, France}
\affiliation{ExoClock Project, \url{https://www.exoclock.space/}}

\author[0000-0003-3759-9080]{Tiffany Kataria}
\affiliation{Jet Propulsion Laboratory, California Institute of Technology, Pasadena, CA 91125 USA}

\author[0000-0002-3205-0147]{Anastasia Kokori}
\affiliation{University College London, Gower Street, London, WC1E 6BT, UK}
\affiliation{ExoClock Project, \url{https://www.exoclock.space/}}

\author[0000-0003-0871-4641]{Harsh Kumar}
\affiliation{Department of Physics, Indian Institute of Technology Bombay, Powai, 400076, India}

\author{Petri Kuossari}
\affiliation{Unistellar Network Citizen Scientist, \url{https://unistellaroptics.com/citizen-science/}}

\author[0000-0003-3559-0840]{Georgios Lekkas} 
\affiliation{Department of Physics, University of Ioannina, Ioannina, 45110, Greece}
\affiliation{Exoplanet Watch, \url{https://exoplanets.nasa.gov/exoplanet-watch/}}

\author[0000-0003-3779-6762]{Alessandro Marchini}
\affiliation{University of Siena, Department of Physical Sciences, Earth and Environment, Astronomical Observatory, Via Roma 56, 53100 Siena, Italy}
\affiliation{ExoClock Project, \url{https://www.exoclock.space/}}

\author[0000-0002-5105-635X]{Nicola Meneghelli}
\affiliation{Unistellar Network Citizen Scientist, \url{https://unistellaroptics.com/citizen-science/}}

\author[0000-0001-8771-7554]{Chow-Choong Ngeow}
\affiliation{Graduate Institute of Astronomy, National Central University, 300 Jhongda Road, 32001 Jhongli, Taiwan}

\author{Michael Primm}
\affiliation{Unistellar Network Citizen Scientist, \url{https://unistellaroptics.com/citizen-science/}}

\author[0000-0003-2167-9764]{Subham Samantaray}
\affiliation{Department of Physics, Indian Institute of Technology Bombay, Powai, 400076, India}

\author{Masao Shimizu (清水正雄)}
\affiliation{Unistellar Network Citizen Scientist, \url{https://unistellaroptics.com/citizen-science/}}

\author{George Silvis}
\affiliation{American Association of Variable Star Observers, 49 Bay State Rd, Cambridge, MA 02138, USA}
\affiliation{Exoplanet Watch, \url{https://exoplanets.nasa.gov/exoplanet-watch/}}

\author{Frank Sienkiewicz}
\affiliation{The Center for Astrophysics, Harvard $\&$ Smithsonian, 60 Garden Street, Cambridge, MA 02138, USA}
\affiliation{Exoplanet Watch, \url{https://exoplanets.nasa.gov/exoplanet-watch/}}

\author[0000-0002-7942-8477]{Vishwajeet Swain}
\affiliation{Department of Physics, Indian Institute of Technology Bombay, Powai, 400076, India} 

\author{Joshua Tan}
\affiliation{Exoplanet Watch, \url{https://exoplanets.nasa.gov/exoplanet-watch/}}

\author{Kalee Tock}
\affiliation{Stanford Online High School, Academy Hall Floor 2 8853, 415 Broadway Redwood City, CA 94063, USA}
\affiliation{Exoplanet Watch, \url{https://exoplanets.nasa.gov/exoplanet-watch/}}

\author[0000-0002-4309-6343]{Kevin Wagner}
\altaffiliation{NASA Hubble Fellowship Program - Sagan Fellow}
\affiliation{Unistellar Network Citizen Scientist, \url{https://unistellaroptics.com/citizen-science/}}
\affiliation{Department of Astronomy and Steward Observatory, University of Arizona, Tucson, AZ 85721 USA}

\author{Ana{\"e}l W{\"u}nsche}
\affiliation{Observatoire des Baronnies Proven{\c c}ales, Route de Nyons, F-05150 Moydans, France}
\affiliation{ExoClock Project, \url{https://www.exoclock.space/}}







\date{June 2022}
\section{Abstract}
The transiting planet HD~80606~b undergoes a 1000-fold increase in insolation during its 111 day orbit due to it being highly eccentric ($e$=0.93). The planet's effective temperature increases from 400~K to over 1400~K in a few hours as it makes a rapid passage to within 0.03~AU of its host star during periapsis. Spectroscopic observations during the eclipse (which is conveniently oriented a few hours before periapsis) of HD~80606~b with the James Webb Space Telescope (JWST) are poised to exploit this highly variable environment to study a wide variety of atmospheric properties, including composition, chemical and dynamical timescales, and large scale atmospheric motions. Critical to planning and interpreting these observations is an accurate knowledge of the planet's orbit. We report on observations of two full-transit events: 7 February 2020 as observed by the TESS spacecraft and  7--8 December 2021 as observed with a world-wide network of small telescopes. We also report new radial velocity observations which when analyzed with a coupled model to the transits greatly improves the planet's orbital ephemeris. Our new orbit solution reduces the uncertainty in the transit and eclipse timing of the JWST era from tens of minutes to a few minutes. When combined with the planned JWST observations, this new precision may be adequate to look for non-Keplerian effects in the orbit of HD~80606~b.

\section{Introduction}

For many years HD~80606~b held the record for the  most highly eccentric planet. Discovered by the radial velocity (RV) technique in 2001 \citep{Naef2001} HD~80606~b has a mass of 4.1~\mj, an orbital period of 111.4~days and an eccentricity of $\epsilon$=0.93. Its eccentricity is currently exceeded only by HD~20782~b with an eccentricity of $\epsilon$=0.95 \citep{Jones2006}. 
HD~80606~b continues to be compelling for further study as it was discovered by Spitzer using the eclipse in early 2009 \citep{Laughlin2009}. The transit was then discovered and announced near-simultaneously in late February 2009 by \cite{Fossey2009}, \cite{Garcia2009}, and by \cite{Moutou2009}. HD 80606 b passes within 0.03~AU  of its host G5V star, during its rapid periastron passage of a few tens of hours, the insolation and temperature of the planet increase dramatically, from 1$\times$ to almost 1000$\times$ Earth-Equivalent and from 400~K to over 1400~K.

These rapid changes, coupled with the fact that HD~806060~b transits and also eclipses (passes behind the star), provide a unique opportunity to explore the dynamical response  of an atmosphere under an extreme external forcing function. Spitzer's photometric observations of eclipses in 2009 and 2010 at 8.0 and 4.5~\mum\, respectively, were used to infer timescales for radiative, dynamical, and chemical processes \citep{dewit2016, Lewis2017}. As noted by \citet{Lewis2017}, ``The time-variable forcing experienced by
exoplanets on eccentric orbits provides a unique and important window on radiative, dynamical, and
chemical processes in planetary atmospheres and an important link between exoplanet observations
and theory." 

The James Webb Space Telescope (JWST) will expand  these studies  dramatically using spectroscopy. Kataria et al.\footnote{Approved Cycle 1 program \#2008. ``A Blast From the Past: A Spectroscopic look at the Flash Heating of HD~80606~b" https://www.stsci.edu/jwst/science-execution/program-information.html?id=2008} will use the MIRI Low Resolution Spectrometer (MIRI/LRS) to observe an eclipse of HD~80606~b from 5--14~\mum\ at a spectral resolution of $\sim$100. Sikora et al\footnote{Approved Cycle 1 program \#2488. ``Real Time Exoplanet Meteorology: Direct Measurement of Cloud Dynamics on the High-Eccentricity Hot Jupiter HD~80606 b" https://www.stsci.edu/jwst/science-execution/program-information.html?id=2488} will explore the formation and evolution of atmospheric clouds at shorter wavelengths using NIRSpec at 2.87-5.18 \mum\ with a resolution of $\sim$2700 to observe the eclipse and periastron passage. These spectral regions contain a wealth of molecular features whose variation will reveal new insights into the chemistry and dynamics of the atmospheres of giant planets.

A challenge to transit and eclipse observations is the gradual erosion of our knowledge of a planet's orbital properties. Uncertainties in the timing of transits and eclipses lead to observing inefficiencies as longer durations must to scheduled to avoid missing some or all of an event \citep[e.g.,][]{Dragomir2020, Zellem2020}. This problem is exacerbated in the case of HD~80606~b where the relevant observations are over a decade old and uncertainties on the eclipse prediction grow with each orbit ($\sim$3 per year). Of particular importance is the knowledge of the time of periastron passage relative to the eclipse as this is needed to link the spectral observations to the insolation profile. 

It was to remedy this growing uncertainty in our knowledge of the ephemerides of HD~80606~b that we undertook to analyze the TESS data and to obtain observations of the transit occurring on 7/8-Dec-2021 (Table~\ref{tab:nominal} and Figure~\ref{fig:Map}) from the ground. We also obtained new RV measurements around the time of periastron to continue to refine the RV solution. Section~$\S$\ref{sec:trans} describes the observations of the transit and  $\S$\ref{sec:PRV} the RV observations. Section~$\S$\ref{sec:analysis} describes the analysis of the various datesets while $\S$\ref{sec:params} uses the combined transit and RV measurements to refine the ephemeris of HD~80606~b and to predict the times of occurrence of future transits and eclipses. 

\begin{figure*}
\centering
\includegraphics[scale=0.4]{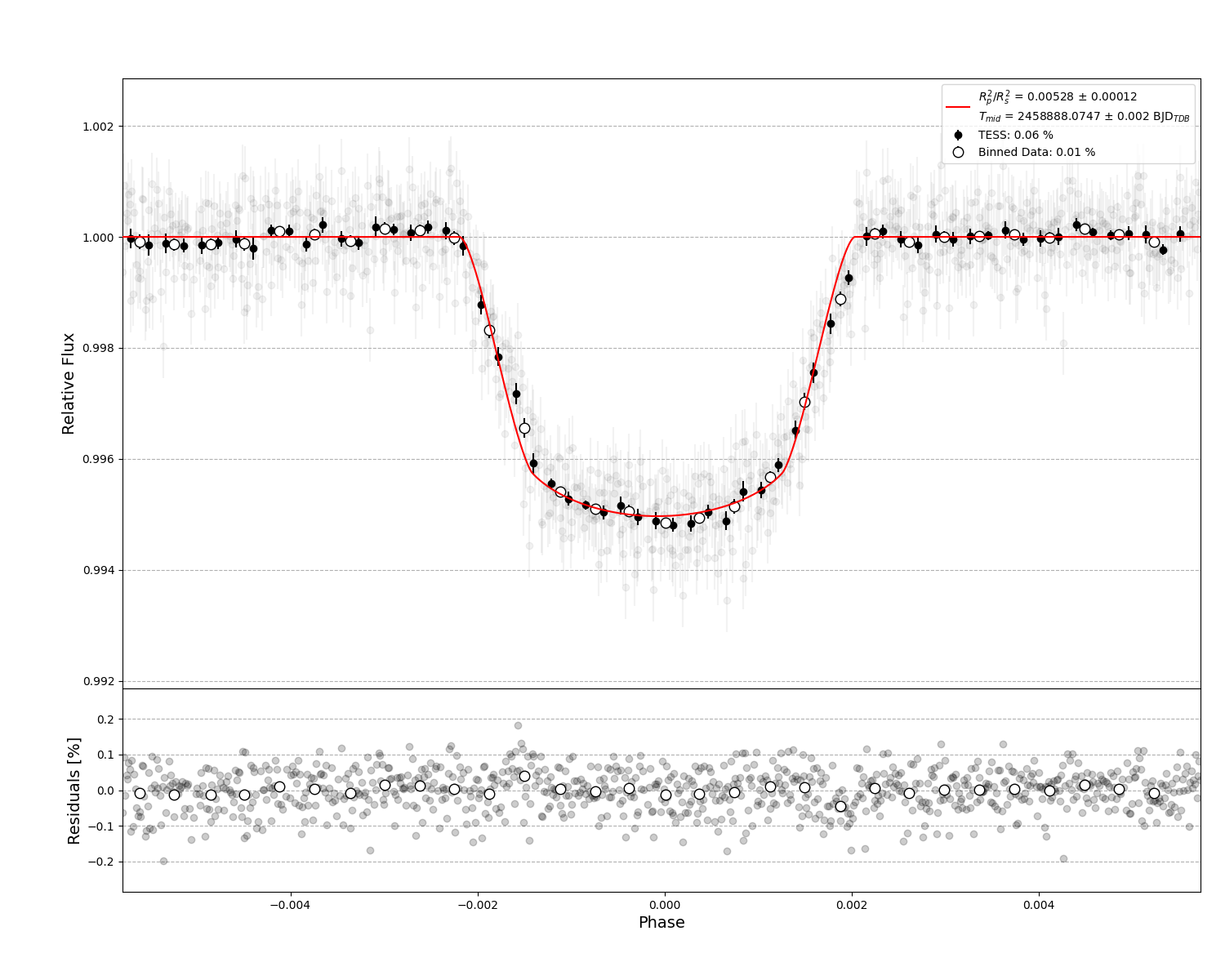}
\caption{ A transit light curve of HD 80606 b measured with the TESS spacecraft using data from Sector 21. The TESS light curve is contaminated with light from a neighboring star causing the transit depth to appear smaller (by about $\sim$48$\%$) than it really is. The plate scale of TESS is $\sim$21'' $\times$ 21'' and that is also coincidentally the distance between the nearby stellar companion, HD~80607, and HD~80606. Light contamination from the roughly equal brightness companion was summed in the aperture used for TESS photometry and will contribute to a smaller measured depth than observations from platforms with a higher imaging resolution, where the light sources can be treated separately. Despite the contamination shrinking the measured depth, we can still detect it to $\sim$44$\sigma$ which is enough to constrain the time of mid-transit to within $\sim$3 minutes. The binned data is purely for visualization purposes and is at two different cadences, 30-minutes in black and 60-minutes in white with a black outline while the transparent points are the original data.
\label{fig:joint_transit}}
\end{figure*}

\begin{deluxetable}{lcl}[t!]
\tablecaption{Orbital Prior for HD~80606~b\label{tab:nominal}}
\tablehead{
\colhead{Parameter} & \colhead{Value} & \colhead{Reference}}
\startdata
T$_{mid}$ (MJD)&2455210.6428$\pm$0.001 &\citet{Bonomo2017}\\
E$_{mid}$ (MJD)&2454424.736 $\pm$0.003 &\citet{Laughlin2009}\\
& 14-Jan-2010 0326 UTC&\\
Period (d)&111.43670$\pm$0.0004 &\citet{Bonomo2017}\\
Eccentricity ($e$)&0.93226$\pm$0.00066&\citet{Bonomo2017}\\
Arg. Periapsis ($\omega_{peri})$&58.97$\pm$0.2 (deg)&\citet{Bonomo2017}\\
&-1.0292$\pm$0.0035 (rad)&\\
Transit Duration (hr)&11.64$\pm$0.25&\citet{Winn2009}\\
\multicolumn{3}{l}{\textit{Prediction for Dec. 2021}}\\
Accum. Unc. (hr)$^1$& 0.4 for  Observed transit\\
T$_{mid}$ (MJD)&2459556.674$\pm$0.016  d &\\
Observed event&08-12-2021 0411 UTC &\\
\enddata
\tablecomments{$^1$Accumulated uncertainty in the timing of the transit occurring  $N_{per}=39$ periods after the reference time, $T_c$. $\sigma T=\sqrt{\sigma(T_c)^2+N_{per}^2\sigma(Period)^2}$ (Eqn.~3 in \citet{Zellem2020})}
\end{deluxetable}

\section{Observations \label{sec:obs}}

A majority of the transit observations for HD 80606 b originate almost a decade ago when it was a targeted by the Spitzer Space Telescope. Since then, there hasn't been a full transit observation in $\sim$10 years although the star has been monitored by radial velocity surveys. In preparation for JWST observations we have combined observations of the 2020 transit taken by TESS with 2021 observations taken from the ground by the Exoplanet Watch program. Finally, the light curve measurements are combined with new and archival radial velocity measurements in order to constrain the orbit parameters and to improve our knowledge of transit and eclipse events over the next decade.

\subsection{2020 Transit With TESS}
The photometric data from TESS were processed using a custom pipeline leveraging optimal aperture selection, systematic detrending with a weighted spline and outlier rejection in order to improve and minimize the scatter in the light curve \citep{Pearson2019b}. The custom pipeline uses multiple aperture sizes during the photometric extraction in order to minimize the scatter in the residuals after fitting a light curve model. Detrending the time series and minimizing scatter in the residuals has been shown to improve light curve quality compared to the default produced from the Science Processing Operations Center (SPOC) pipeline \citep{Jenkins2016} which is based on the Kepler mission pipeline \citep{Jenkins2010}.

TESS is capable of high precision measurements for this system due to the host star being bright (V=9.0 mag). However, TESS's large pixel size (21\arcsec) is less than ideal for HD~80606 due to the presence of HD~80607, a nearby  companion of similar spectral type and brightness (V=9.07 mag) separated by 20.5\arcsec. Stellar blends  dilute the transit signal causing a larger planet to mistakenly appear smaller \citep[e.g.,][]{Ciardi15, Zellem2020}. In the reduction of TESS data, a wide aperture was used and includes light from both stars. Therefore, our estimate for the transit depth is underestimated. The estimated contamination is around $\sim$48$\%$ and translates to a corrected transit depth $\sim2\times$ greater than what we directly measure. Despite the contamination decreasing the transit depth, we still detect the transit at over 40 $\sigma$ which allows for a strong constraint on the time of mid-transit to within a few minutes (see Table~\ref{tab:newmid} and Figure \ref{fig:joint_transit}).

\begin{deluxetable*}{lllllll}
\centering
\tablecaption{Transit Observing Facilities\label{tab:facilities}}
\tablehead{
\colhead{Facility} & \colhead{Location (N,E)} & \colhead{Size (m)}& \colhead{UTC Start (Phase)} & \colhead{UTC Stop (Phase)}& \colhead{Precision \% $^{1}$ }& \colhead{N. Images} }
\startdata
Transiting Exoplanet  & Space & 0.1 & 2020-02-07 20:32:00 (-0.0054) & 2020-02-07  07:06:00 (0.0054) & 0.06 & 1520 \\ 
Survey Satellite (TESS) & &&&&\\
\hline
Exoplanet Watch [HJEB]  & (30.7, -104.2) & 0.4       & 2021-12-06 08:21:36 (-0.0166) &2021-12-06 09:40:50 (-0.0161)  & 1.31 & 225 \\ 
Las Cumbres (LCO) & (30.7, -104.2) & 0.4             & 2021-12-07 06:48:56 (-0.0079) & 2021-12-07 07:39:54 (-0.0082) & 1.26 & 218 \\
Las Cumbres (LCO) & (30.7, -104.2) & 0.4             & 2021-12-07 09:46:56 (-0.0068) & 2021-12-07 10:38:05 (-0.0071) & 0.77 & 225 \\
Las Cumbres (LCO) & (30.7, -104.2) & 0.4             & 2021-12-07 11:35:45 (-0.0064)  & 2021-12-07 12:26:18 (-0.0061) & 1.21 & 221 \\
Exoplanet Watch [NCC] & (23.5, 120.9)  & 0.4         & 2021-12-07 17:34:11  (-0.0042) & 2021-12-07 20:13:20 (-0.0032) & 1.01 & 481 \\ 
GROWTH-India & (32.8, 79.0)  & 0.7                   & 2021-12-07 19:52:49 (-0.0033) & 2021-12-08 00:40:41 (-0.0015) & 0.53 & 609 \\ 
Unistellar eVscope 2 (2rz)  & (49.2, -0.4)  & 0.11   & 2021-12-07 20:49:47 (-0.0030) & 2021-12-08 01:38:22 (-0.0012) & 1.09 & 126 \\
Unistellar eVscope (etx) & (49.2, -0.4) & 0.11       & 2021-12-07 20:48:29 (-0.0030) & 2021-12-08 01:37:27 (-0.0012) & 0.63 & 131 \\
Unistellar eVscope (257) & (60.8, 24.4) & 0.11       & 2021-12-07 21:41:31 (-0.0027) & 2021-12-08 00:17:56 (-0.0017) & 0.36 & 79 \\
Unistellar eVscope (3mh) & (45.3, 11.1)  & 0.11      & 2021-12-07 22:24:41 (-0.0024) & 2021-12-08 01:41:27 (-0.0012) & 0.67 & 55 \\
Exoplanet Watch [GDAI]  & (39.0, -108.2) & 0.4       & 2021-12-08 03:37:37 (-0.0004) & 2021-12-08 11:46:49 (0.0026)  & 3.11 & 503 \\
Unistellar eVscope (rev) & (30.4, 97.8) & 0.11       & 2021-12-08 04:26:52 (-0.0001) & 2021-12-08 08:09:55 (0.0013)  & 0.50 & 101 \\
Unistellar eVscope (sdp) & (32.2, -111) & 0.11       & 2021-12-08 05:17:14 (0.0002)  & 2021-12-08 12:18:15 (0.0028)  & 0.78 & 155 \\
Exoplanet Watch [RJBA]  & (34.1, -118.1) & 0.15      & 2021-12-08 06:09:47 (0.0005)  & 2021-12-08 12:08:50 (0.0027)  & 1.47 & 569 \\ 
Las Cumbres (LCO) & (30.7, -104.2) & 1               & 2021-12-08 06:41:20 (0.0007)  & 2021-12-08 12:17:36 (0.0028)  & 0.33 & 391 \\
Exoplanet Watch [HJEB]  & (30.7, -104.2) & 0.4       & 2021-12-08 06:46:01 (0.0007)  & 2021-12-08 07:36:43 (0.001)   & 1.29 & 225 \\
Las Cumbres (LCO) & (30.7, -104.2) & 0.4             & 2021-12-08 11:35:50 (0.0025)  & 2021-12-08 12:26:33 (0.0029)  & 0.80 & 225 \\
Unistellar eVscope (8cm) & (35.1, 134.4) & 0.11      & 2021-12-08 13:19:08 (0.0032) & 2021-12-08 14:14:42 (0.0035) & 1.54 & 26 \\
Exoplanet Watch [NCC] & (23.5, 120.9)  & 0.4         & 2021-12-08 16:04:28 (0.0042) & 2021-12-08 20:08:09 (0.0057) & 0.80 & 516 \\ 
Unistellar eVscope 2 (2rzB) & (49.2, -0.4) & 0.11    & 2021-12-08 21:47:08 (0.0063) & 2021-12-08 23:47:48 (0.0071) & 1.08 & 88 \\
Unistellar eVscope (etxB) & (49.2, -0.4) & 0.11      & 2021-12-08 21:48:00 (0.0064) & 2021-12-08 23:39:20 (0.007) & 1.25 & 152\\
Exoplanet Watch [BARO] & (32.6, -116.3)  & 0.43 & 2021-12-09 01:26:11 (0.0077) & 2021-12-09 01:55:10 (0.0079) & 0.97 & 98 \\ 
Exoplanet Watch [LGEC]  & (28.3, -16.6)  & 0.4      & 2021-12-09T02:06:25 (0.008)  & 2021-12-09 02:15:10 (0.008) & 0.80 & 29 \\ 
Exoplanet Watch [FMAA]  & (31.7, -111.1) & 0.15     & 2021-12-09T04:41:25 (0.009)  & 2021-12-09 12:06:02 (0.012) & 1.79 & 130 \\ 
\enddata
\tablecomments{$^1$Standard deviation of the residuals \\The observations are split between the archival measurements (top) and those taken for the same transit (bottom).\\For the exoplanet watch observations the letters in brackets represent the AAVSO Observer code so the datasets can be easily referenced in the future and searchable on their archive. }
\end{deluxetable*}

\begin{figure}[b!]
\centering
\includegraphics[width=0.5\textwidth]{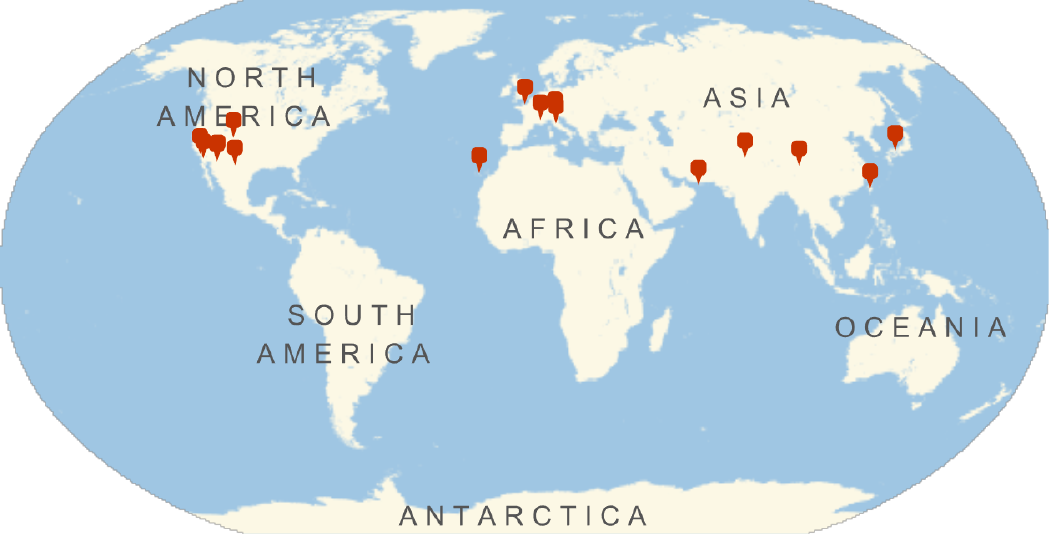} \\
\caption{A map of the facilities in the global network of small telescopes used to observe the transit on 2021, Dec 7/8.
\label{fig:Map}}
\end{figure}

\subsection{2021 Transit from the Ground}\label{sec:trans}

HD~80606~b's long transit duration, over 11.5~hr \citep{Pont2009,Winn2009},  and the accumulated uncertainty in its time of occurrence, make a world-wide program of coordinated observations essential. Fortunately, networks of small and modest sized telescopes (e.g., Exoplanet Watch\footnote{https://exoplanets.nasa.gov/exoplanet-watch/}, ExoClock\footnote{http://exoclock.space}, Unistellar\footnote{https://unistellaroptics.com/}) are now in place to support programs of this type. The global observational campaign to measure the 2021 December 7--8 transit of HD~80606~b presented here was  coordinated by Exoplanet Watch. The various observatories that contributed a transit measurement in December are shown in Figure \ref{fig:Map}.

\subsubsection{Exoplanet Watch}
Exoplanet Watch is a citizen science project funded by NASA's Universe of Learning\footnote{https://www.universe-of-learning.org} for observing exoplanets with small, ground-based telescopes to maintain ephemerides and to ensure the efficient use of large telescopes, discover new exoplanets via transit timing variations, resolve blended pairs, monitor for stellar variability, and confirm exoplanet candidates \citep{Zellem2019, Zellem2020}. Anyone is able to contribute observations to a public data archive\footnote{https://app.aavso.org/exosite/}, hosted by the American Association of Variable Star Observers\footnote{http://aavso.org}, where they are analyzed on a regular basis and used to refine exoplanet ephemerides\footnote{https://exoplanets.nasa.gov/exoplanet-watch/results/}. The observations listed under Exoplanet Watch in Table~\ref{tab:facilities} are currently available online and are linked to their AAVSO observer code. A majority of the users contributed at least one hour of observations using telescopes smaller than 0.5-meters. A few notable contributors to the network include the Boyce-Astro Research Observatory (BARO) located at an observing site near Tierra Del Sol and Campo, California. BARO includes a 17-inch telescope and a ZWO ASI 1600 CMOS camera. The observing configuration provides a 8.3'$\times$6.3' field of view with a plate scale of 0.107'' per pixel. Additionally, an individual user was able to capture part of transit egress from the top of the Cahill building on the campus of California Institute of Technology using a 6~inch telescope and the ASI 224MC camera. 

Another contributor is the  MicroObservatory which hosts a network of automated remote reflecting telescopes, each with a 6-inch mirror, 560-mm focal length, and KAF1402ME CCD with 6.8-micron-sized pixels. With 2×2 pixel binning, the image size is 650×500 pixels  at a pixel scale of approximate 5”/px. MicroObservatory takes images of exoplanet systems daily and makes the images publicly available for educational use via their DIY Planet Search program\footnote{https://mo-www.cfa.harvard.edu/MicroObservatory/}.

\begin{figure*}[b!]%
\centering
\includegraphics[width=0.99\textwidth]{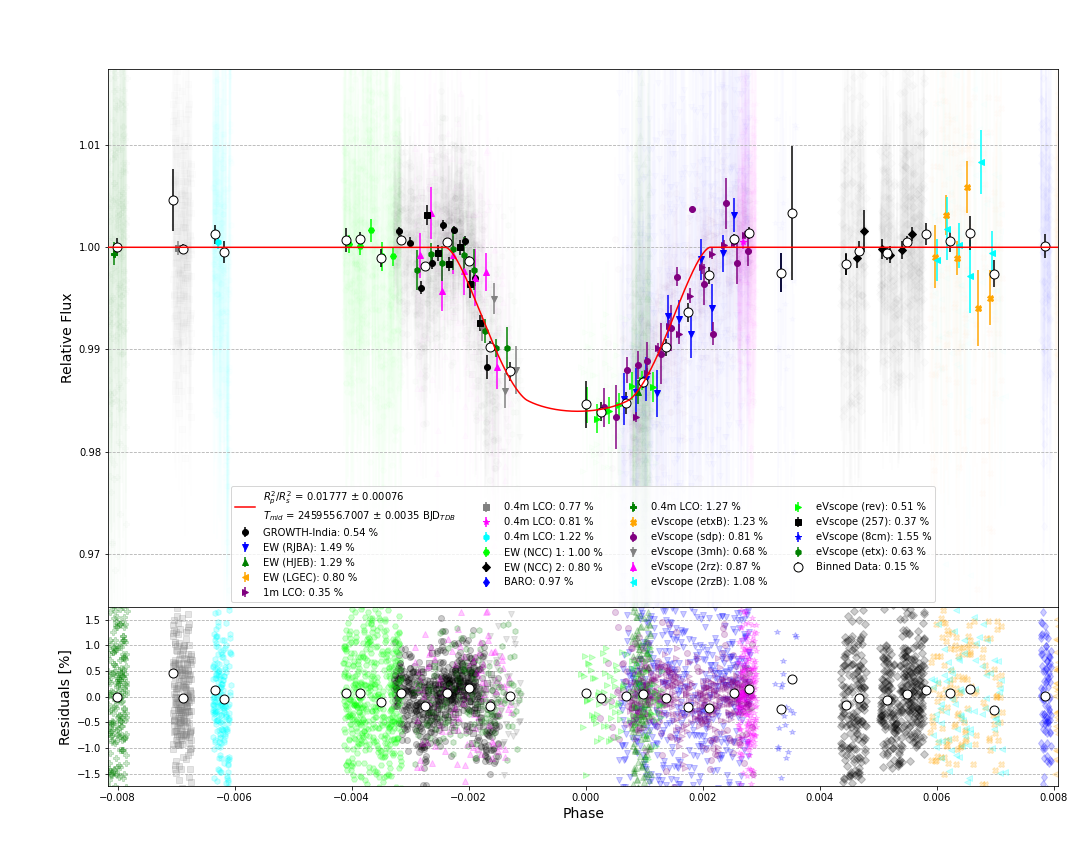} \\
\caption{\textit{Top:} The combined light curve showing the complete transit of HD~80606~b on 7--8 Dec 2021 along with a model fit to the observations (red line). The data are binned to a resolution of 30~minutes for each individual data set and 60~minutes for the combined data set (empty circles) for the purposes of  visualization. Each observation is fit simultaneously with equation \ref{expam} and requires a separate airmass model for detrending. A mosaic of individual light curves can be found in the appendix (see Figure~\ref{fig:mosaic}). \textit{Bottom:} Residuals for the light curve model are displayed at the native resolution except for a binned version shown in white circles. The standard deviation of the residual scatter is reported in the legend on the top subplot. 
\label{fig:BestFit}}
\end{figure*}

\subsubsection{LCO Network}
Las Cumbres Observatory (LCO) is a global telescope network consisting of multiple meter and sub-meter sized telescopes at various locations around the Earth. HD~80606 was observed over the course of 3 days from multiple locations in the LCO network. Unfortunately,   weather  clouded-out most of the Northern Hemisphere so that only a few sites acquired data. A majority of the usable observations come from LCO's telescopes at McDonald Observatory in Texas and Teide Observatory in Tenerife. LCO's 0.4-meter telescopes contain SBIG CCD cameras with a field of view  $\sim$29' $\times$ 29', corresponding to a plate scale of 0.571''/pixel. The 1-meter telescope apart of LCO contains a Sinistro imager with a 26' $\times$ 26' field of view and a plate scale of 0.39''/px. All of the LCO observations were acquired with the R filter and some observatory-specific details are highlighted in Table~\ref{tab:facilities}.

\subsubsection{Unistellar Network}

The Unistellar Network is a global community of citizen scientist observers with Unistellar telescopes who have open access to observing campaigns organized by SETI Institute astronomers, including exoplanet transit observations. Seven different eVscopes (``Enhanced Vision Telescopes'') acquired nine observations of HD~80606~b from six different observing locations in North America, Europe, and Japan (Table~\ref{tab:facilities}). Of those observations, seven were collected using the Unistellar eVscope~1, which is a 4.5-inch reflecting telescope with a Sony IMX224LQR CMOS sensor at its prime focus. The camera's field of view is 37.0$\arcmin$ x 27.7$\arcmin$ with a plate scale of 1.7 $\arcsec$/pixel. Individual images had an exposure time of 3.970~s and sensor gain of 2 dB. The two remaining observations were collected using the Unistellar eVscope~2, which shares the design of the eVscope~1 but has a Sony IMX347LQR CMOS sensor. The camera's field of view is 45.3$\arcmin$ x 34.0$\arcmin$ with a plate scale of 1.3 $\arcsec$/pixel. Individual images had an exposure time of 3.970~s and sensor gain of 0~dB (no digital gain).

\subsubsection{ExoClock Project}

In addition to the TESS and December transit of HD 80606 b we also report on three additional transit measurements from the project ExoClock \citep{Kokori2021}. The ExoClock project is an open-access citizen science project aimed at conducting transit measurements of exoplanets targeted by the Ariel Mission \citep{Tinetti16}. The three measurements were taken from ground-based observatories in Europe with mid-transit measurements reported in Table~\ref{tab:oldmid}.

\subsubsection{GROWTH}

The Global Relay of Observatories Watching Transients Happen (GROWTH) network involves over a dozen institutions dedicated to the follow-up of transient events \citep{Kasliwal2019}. Among these, a number of Asian observatories within the GROWTH collaboration participated in the 2021 Dec 7/8 campaign, providing critical data during transit ingress. The GROWTH-India Telescope (GIT) is a 0.7m fully robotic telescope located at the Indian Astronomical Observatory (IAO), Hanle-Ladakh. The telescope is equipped with an Andor Ikon230XL CCD camera which provides a Field of view of $\sim0.5~\rm{deg}^2$. GIT observed HD~80606~b for $\sim 5$~hrs on night of Dec~7, 2022, obtaining a total of 609 images. The details of the observations are provided in  Table~\ref{tab:facilities}. Data were reduced following standard procedures, and photometry was performed with EXOTIC as described in 3.2.

\subsection{Transit Data Reduction}

Data reduction and calibrations of the individual science images was done by each observer or their group. We encouraged all groups to acquire at least a bias and flat-field frame in order to reduce noise and normalize pixel to pixel changes in sensitivity, respectively. We provided an open-source package for aperture photometry and light curve fitting in order to make extracting the time series easy and optimal with respect to minimizing sources of noise. The EXOplanet Transit Interpretation Code\footnote{https://github.com/rzellem/EXOTIC} (EXOTIC; \citealt{Zellem2020}; Fatahi et al. \textit{in prep.}) can calibrate images (i.e. bias, flat and dark), plate solve images for better centroiding and conducts an optimization over comparison star selection and aperture when extracting the photometric timeseries. After conducting aperture photometry, all of the time series files were combined in order to produce the global light curve shown in Figure \ref{fig:BestFit}. A mosaic of the individual observations is shown in the appendix (see Figure \ref{fig:mosaic}.)

\begin{deluxetable}{lll}
\centering
\tablecaption{Archival Ephemeris Times\label{tab:oldmid}}
\tablehead{
\colhead{BJD$_{TBD}$} & \colhead{Reference} & \colhead{Status}}
\startdata
    2454424.736 $\pm$ 0.003 & \cite{Laughlin2009} & Full Eclipse \\
    2454876.316 $\pm$0.023  & \cite{Pont2009} & Partial Transit \\
    2454876.338 $\pm$ 0.017 & \cite{Kokori2021} & Partial Transit\\ 
    2454987.7842 $\pm$0.0049 & \cite{Winn2009} & Full Transit\\
    2455099.196 $\pm$ 0.026  & \cite{Shporer2010} & Partial Transit\\
    2455210.6420 $\pm$0.001  & \cite{Hebrard2010} & Full Transit \\
    2455210.6502 $\pm$ 0.0064 & \cite{Shporer2010} & Full Transit\\
    2457439.401 $\pm$ 0.012 & \cite{Kokori2021} & Partial Transit\\ 
    2459222.401 $\pm$ 0.016 & \cite{Kokori2021} & Partial Transit\\  
\enddata
\end{deluxetable}

\begin{deluxetable}{ll}
\centering
\tablecaption{New Mid-transit Times\label{tab:newmid}}
\tablehead{
\colhead{Facility} & \colhead{BJD$_{TBD}$}}
\startdata
TESS & 2458888.07466 $\pm$ 0.00204 \\
Multiple (7--8 Dec. 2021) & 2459556.7007 $\pm$ 0.0035 \\
\enddata
\end{deluxetable}

\begin{deluxetable}{lll}
\centering
\tablecaption{New Radial Velocity Observations\label{tab:NEW_RV}}
\tablehead{
\colhead{Instrument} & \colhead{BJD$_{TBD}$}& \colhead{Relative RV}}
\startdata
HIRES &2459514.0886&-133.668$\pm$1.168\\
APF & 2459533.0674 & 37.779$\pm$2.332\\
APF & 2459535.9405 & 15.584$\pm$8.951\\
APF & 2459541.0692 & -13.924$\pm$2.248\\
APF & 2459541.8002 & -9.460$\pm$2.288\\
APF & 2459544.0027 & -28.552$\pm$2.413\\
\enddata
\end{deluxetable}

\begin{deluxetable}{lll}
\centering
\tablecaption{Archival Radial Velocity Observations\label{tab:OLD_RV}}
\tablehead{
\colhead{Instrument} & \colhead{BJD$_{TBD}$}& \colhead{Relative RV}}
\startdata
ELODIE & 2452075.359 & -134.46$\pm$13\\
... \\
HIRES$_K$ &2452219.162 & -85.11$\pm$1.6\\
... \\
HRS &2453433.606&119.8$\pm$8.6\\
... \\
HIRISE$_J$ &2453398.854&-171.57$\pm$0.89\\
... \\
SOPHIE &2454876.729&222.1$\pm$5\\
... \\
\enddata
\tablecomments{These measurements are available online in a machine readable format \footnote{\url{https://exofop.ipac.caltech.edu/tess/view_tag.php?tag=418623}}. }
\end{deluxetable}

\subsection{Radial Velocity Observations\label{sec:PRV}}

New radial velocity observations were obtained around periapsis in December 2021 using the Levy spectrometer on the 2.4m Automated Planet Finder telescope (APF) \citep{Vogt2014} and the High Resolution spectrometer (HIRES, on the 10m Keck I telescope). The new RV measurements are processed using standard data reduction techniques described in \citet{Butler1996}. The APF and HIRISE RV values are measured using an Iodine cell-based design in order wavelength calibrate the stellar spectrum. The spectral region from 5000-6200 $\AA$ is used for measuring the radial velocities. The new observations are listed in Table~\ref{tab:NEW_RV}. We used a total of 593 RV measurements spanning 22 years for the data analysis (see Figure \ref{fig:joint_rv}) and they are available in a machine readable format online (see Table \ref{tab:OLD_RV}).


\section{Analysis\label{sec:analysis}}

The newly acquired data of HD~80606 b along with the historical measurements for  RV, transit and eclipse are analyzed in a self-consistent manner in order to place constraints on the system parameters. The radial velocity observations help constrain the orbit and alignment of HD~80606~b, which is particularly important considering the high eccentricity of the planet can drastically change the transit duration based on the argument of periastron \citep{Hebrard2010}. The transit observations help  the size of the planet once the orbit is reliably known and disentangled from degeneracies involving the stellar radius,  inclination and contamination by HD 80607. Additionally, using the measured times of mid-transit and mid-eclipse we can search for deviations from a Keplerian orbit, which is potentially indicative of a companion in the system (\citealt{Holman2005}; \citealt{Nesvorney2008}).

\subsection{Global Light Curve Analysis}

Observations for the transit of HD~80606~b on the night of 2021 December 7--8 are combined and fitted simultaneously in order to derive the time of mid-transit and radius ratio between the planet and star. Since each observation was acquired at a different location, it requires individual treatment of extinction from Earth's atmosphere. We adopt a parameterization \citep[e.g.,][]{Pearson2019a} which scales exponentially with airmass and has resemblance to a solution of the radiative transfer equation when the source function is $I(\tau) = I(0)e^{-\tau}$. The following equation is used to maximize the likelihood of the transit model and airmass signal simultaneously:

\begin{equation} \label{expam}
F_{obs} = a_{0} e^{a_{1} \beta }  F_{transit}.
\end{equation}

\noindent Here $F_{obs}$ is the flux recorded on the detector, $F_{transit}$ is the actual astrophysical signal (i.e., the transit light curve, given by pyLightcurve \citep{Tsiaras2016}, $a_{i}$ are airmass correction coefficients and $\beta$ is the airmass value. Since the underlying astrophysical signal is shared between all the observations we leave $R_{p}/R_{s}$ and $T_{mid}$ as free parameters during the retrieval and share the values between each dataset.

The free parameters are optimized using the multimodal nested sampling algorithm called UltraNest (\citealt{Feroz2008}; \citealt{Buchner2014}; \citealt{Buchner2017}). Ultranest is a Bayesian inference tool that uses the Monte Carlo strategy of nested sampling to calculate the Bayesian evidence allowing simultaneous parameter estimation and model selection. A nested sampling algorithm is efficient at probing parameter spaces which could potentially contain multiple modes and pronounced degeneracies in high dimensions; a regime in which the convergence for traditional Markov Chain Monte Carlo (MCMC; e.g., \citealt{ford05}) techniques becomes comparatively slow (\citealt{Skilling2004}; \citealt{Feroz2008}). Convergence for such a large retrieval can take a long time if the priors are very large and sometimes the solutions will not converge at all within a given range for likelihood evaluations for such a large dataset. Therefore, to aid with convergence, each observation was fit individually before being fit simultaneously and given priors to reflect $\pm5\sigma$ around the individual fits. The nested sampling algorithm runs for 500,000 likelihood evaluations before terminating with the resulting posterior distribution shown in Figure \ref{fig:posterior}. An open source version of the global retrieval is available through the EXOTIC repository on GitHub \footnote{https://github.com/rzellem/EXOTIC}. A non-linear 4 parameter limb darkening model is used for the both the ground-based measurements and TESS but corresponding to their respective filters \citep{Morello2020}.

\begin{figure}[h]
\centering
\includegraphics[width=0.5\textwidth]{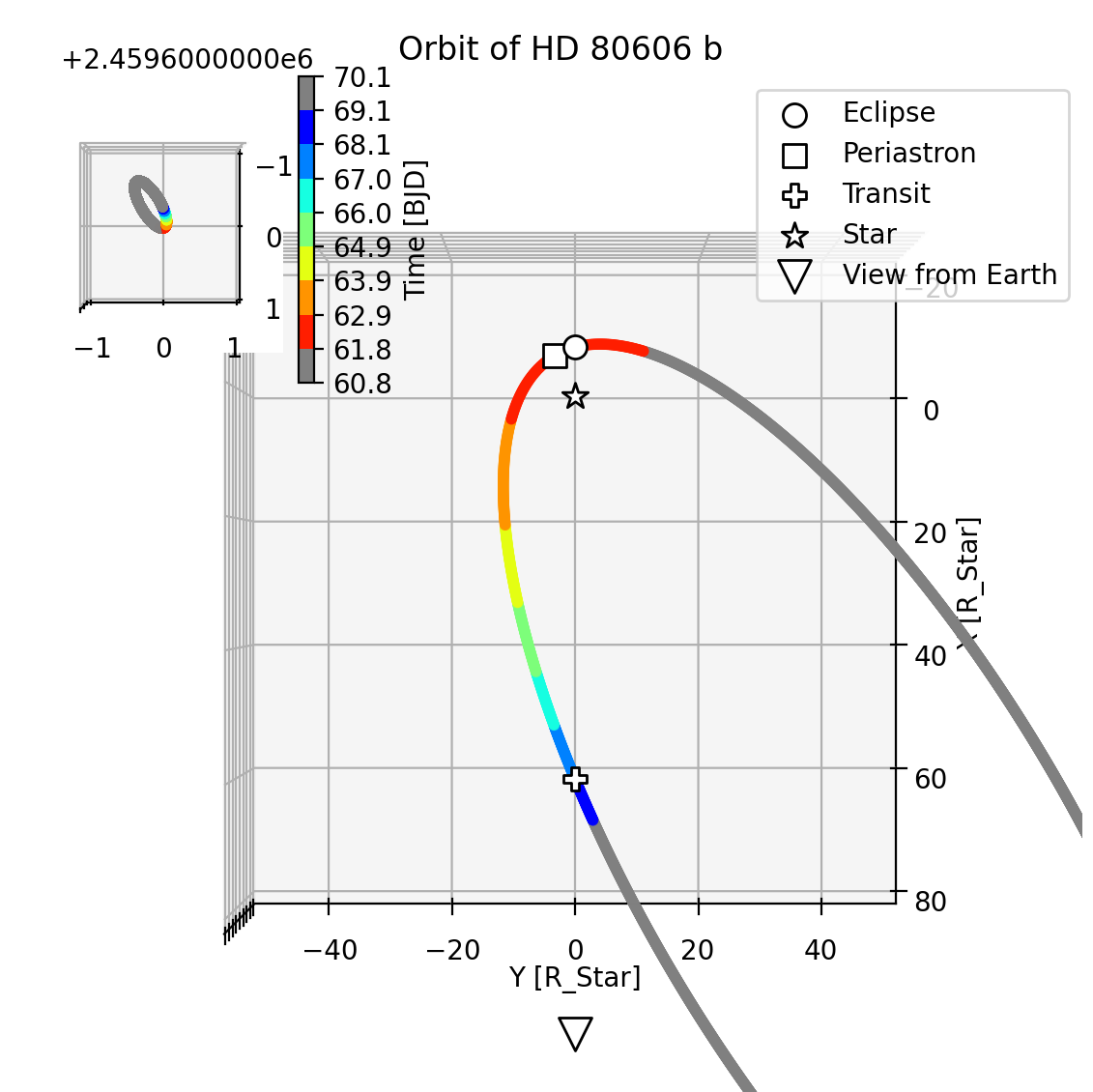}
\caption{Position vectors for the HD~80606 system showing the planet and star plotted over the course of one orbital period for the planet. The colored segments represent chunks of the orbit spanning $\sim$1 day. The big plot has a viewing angle 90 degrees above the line of sight. The small subplot also has a top-down view but of the star's orbit. The markers indicate where mid-transit, mid-eclipse and periastron occur for the planet.
\label{fig:orbit}}
\end{figure}

\subsection{Radial Velocity Analysis}
The archival and new RV measurements (Table~\ref{tab:NEW_RV} and Table~\ref{tab:OLD_RV}) are analyzed using a joint simultaneous fit between a TESS light curve and historical measurements for mid-transit and mid-eclipse in order to constrain a consistent orbital solution across 10 years of heterogeneous data. The radial velocity model uses the same orbit equation and Keplerian solver as the transit light curve model (PyLightcurve; \citealt{Tsiaras2016}). The orbit equation used in the transit model is

\begin{equation}
    r_t = \frac{a}{R_s}\frac{(1-e^2)}{(1+e*cos(\nu_t))}
\end{equation} where $a$ is the semi-major axis, R$_s$ is the stellar radius, $e$ is the eccentricity, and $\nu$ is the true anomaly at some time $t$. The true anomaly can be solved for using equations 1 and 2 in \citealt{Fulton2018} by finding the root of an equation to get the eccentric anomaly which is then used to compute the true anomaly. The orbit equation is projected onto a Cartesian grid which is necessary for the transit model and useful for taking the dot product along our line of sight ensuring it matches the transit geometry (see Figure~\ref{fig:orbit}). The projection along the x-axis, or our line of sight is

\begin{equation}
x_t = r_t sin(\nu_t + \omega) sin(i)
\end{equation}

where $i$ is the inclination of the orbit and $\omega$ is the argument of periastron. The star's velocity is estimated after applying a scaling relation to the planet's orbit assuming it is in a two body system. Coupling the orbit solutions ensures a self consistent system where gravity balances the centripetal acceleration of the planet. The velocity vector of the planet is scaled to match that for the star's orbit and then projected along a line of sight in order to produce the RV signal. A velocity is estimated by evaluating the orbit equation twice in order to compute a numerical derivative using a time step of $\sim$8.5 seconds (0.0001 day):

\begin{equation}
    v_{r,t} = \frac{M_p}{M_s} R_s \frac{x_{t+\Delta t}-x_t}{\Delta t}
\end{equation}

In addition to scaling the planet's orbit by a mass ratio to mimic the stellar position it must also be scaled by the stellar radius in order to acquire units of meters. The stellar radius is given a Gaussian prior during the retrieval process in order to reflect uncertainties on that scale factor and because it is correlated with the planet's inclination. For instance, for a given transit duration there could be a small star with a non-inclined planet or a big star with an inclined planet. Either way they can produce the same transit duration and it is difficult to disentangle the two parameters without an additional constraint on the likelihood function (e.g. some spectral modelling is needed to constrain the stellar properties). We do not have enough information to uniquely constrain the stellar radius and inclination simultaneously which leads to a degeneracy in our retrieval if each parameter uses a uniform prior. Therefore, the stellar radius is given a Gaussian prior which is constructed to be consistent with past derivations in the literature (\citealt{Bonomo2017}; \citealt{Rosenthal2021}).

\begin{figure*}[b!]%
\centering
\includegraphics[width=0.9\textwidth]{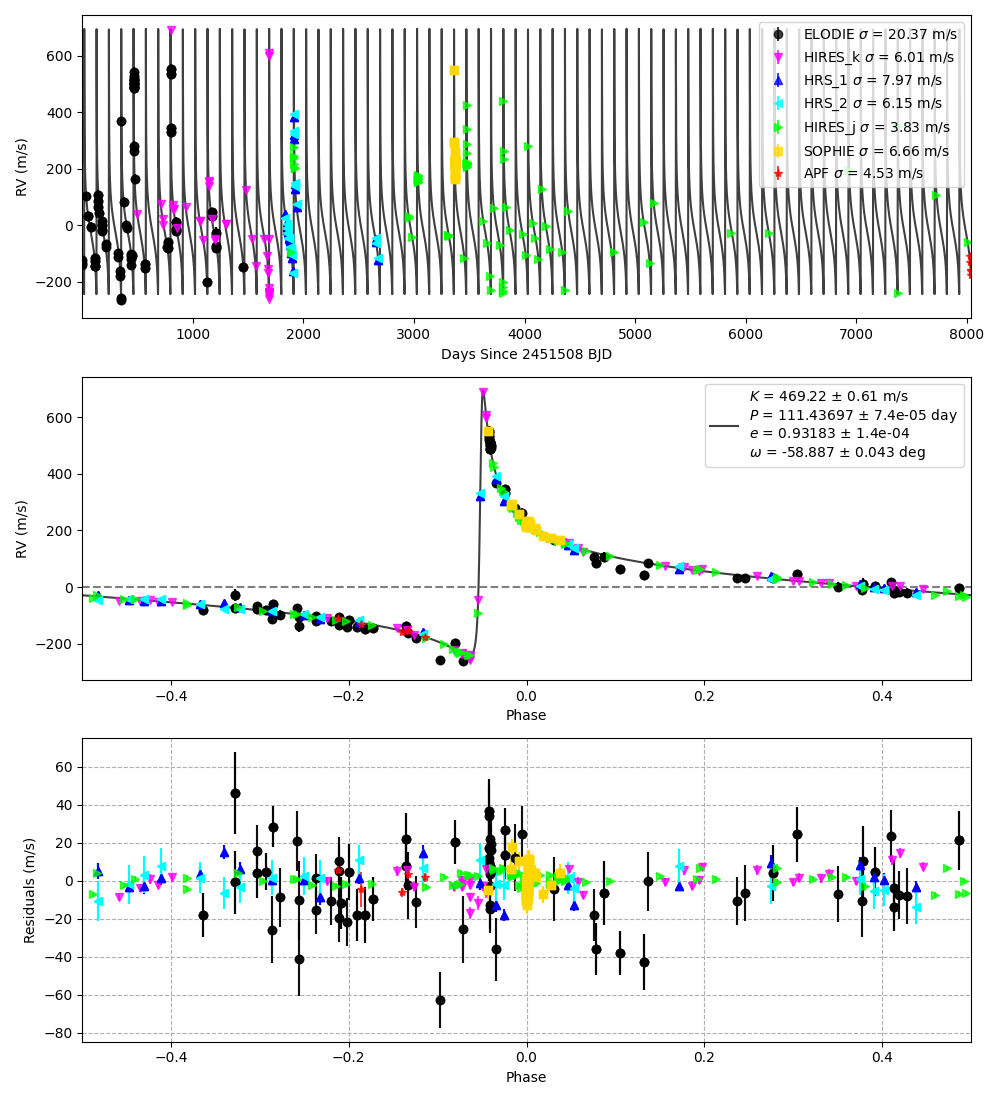}
\caption{Data from 2000 to 2022 show the extremely eccentric orbit of HD~80606~b. The time series RV measurements are plotted in the top panel, the best-fit model is in the middle panel, and the residuals are on the bottom panel. The standard deviation of the residuals is listed in the legend of the top subplot for each dataset. 
\label{fig:joint_rv}}
\end{figure*}



\subsection{Joint simultaneous fit}
Fitting three different types of measurements in a joint analysis requires a likelihood function with contributions from each data set. The system parameters are used to generate a coupled physical model for the transit, RV and ephemeris data in order to enforce consistency between the data sets. The likelihood function includes the sum of the chi-square values when comparing the data sets to their respective model. The TESS light curve is compared to a transit model in a manner similar to the global fit for all the ground-based measurements except the airmass correction is left out. The historic mid-transit and mid-eclipse measurements are compared to a linear ephemeris and then folded into the total chi-squared estimate. The radial velocity measurements are also folded into the total chi-squared however the uncertainties are adjusted prior to the joint fit. The radial velocity likelihood ($\mathcal{L}$) adopts a parameterization similar to RADVEL \citep{Fulton2018} in order to account for underestimated uncertainties,
\begin{equation} \label{rv_likelihood}
   \mathcal{L}_{RV} = -\frac{1}{2} \sum_{i} \sum_{t} (\frac{d_t - v_{r,t}}{\sigma_{i,t} + \sigma_i})^2
\end{equation}
\noindent where d$_t$ is the velocity measurement at time, $t$, $v_{r,t}$ is the Keplerian model predicted for each RV measurement, $\sigma_{i,t}$ is the original uncertainty on the radial velocity measurement and $\sigma_i$ is an RV jitter term for each data set, $i$. The jitter term is set after an individual fit to the radial velocity data and before the joint fit. The jitter term scales the uncertainty such that the average uncertainty is roughly equal to the standard deviation of the residuals from the individual fit. Additionally, the solution to the individual fit $\pm$ 5 sigma is used to constrain the priors for the joint fit. Our uncertainty scaling is similar to RADVEL however we do not include a penalty term which is required when fitting for an error scaling term. We adopt an easier correction for underestimated uncertainties while still being able to leverage the optimizations behind nested sampling. The errors are scaled after an individual fit to the RV data such that the average uncertainty is roughly equal to the scatter in the residuals for that particular data set. After inflating each uncertainty, we found our error estimate for orbital period increased by a factor of $\sim$2 and other orbit parameters similarly. 


The likelihood function for the joint fit has contributions from transit data, RV measurements and historic ephemerides using
\begin{equation} \label{joint_likelihood}
   \mathcal{L}_{Joint} = \mathcal{L}_{RV} + \mathcal{L}_{Transit} + \mathcal{L}_{Mid-transit} + \mathcal{L}_{Mid-eclipse}
\end{equation}. 
The likelihood function for mid-transit and mid-eclipse represent the error for a linear ephemeris estimate compared to existing measurements  Whereas the transit likelihood function uses the photometric time-series.  Nested sampling is used to efficiently explore a large parameter space defining the system and to build a posterior distribution with which to infer uncertainties \citep{Buchner2021}. The free parameters include orbital period, time of mid-transit, inclination, argument of periastron, eccentricity, a planet mass and the radius ratio between the planet and star. Posteriors for the free parameters in the joint fit are shown in Figure 9. We also include a Gaussian prior on the stellar radius because it is needed to convert our radial velocity model into meters. The stellar radius is degenerate with inclination and difficult to constrain if left as a uniform prior. Another relationship in the posteriors is the perfect correlation between eccentricity and argument of periastron. We have seen similar correlations when fitting for $a_{0}$ and $\gamma$ that allowed us to simplify the retrieval and solve for them instead. It is theoretically possible to remove one of these parameters ($e$ or $\omega$) from the sampling process and solve for the other at run-time without having to build it into the posteriors. That solution however requires solving a transcendental equation on top of the existing orbit solution and would increase the computation time of the likelihood function. Therefore, we include both $e$ and $\omega$ in the retrieval and let the sampler handle the correlation which decreases its efficiency slightly.

\begin{figure*}
\centering
\includegraphics[width=0.49\textwidth]{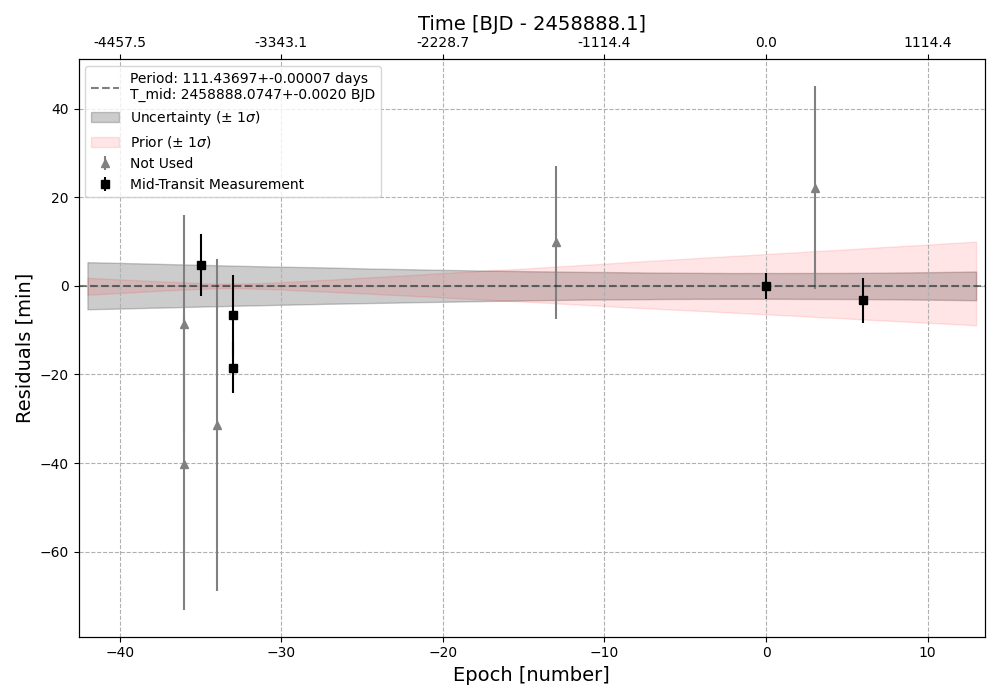}
\includegraphics[width=0.49\textwidth]{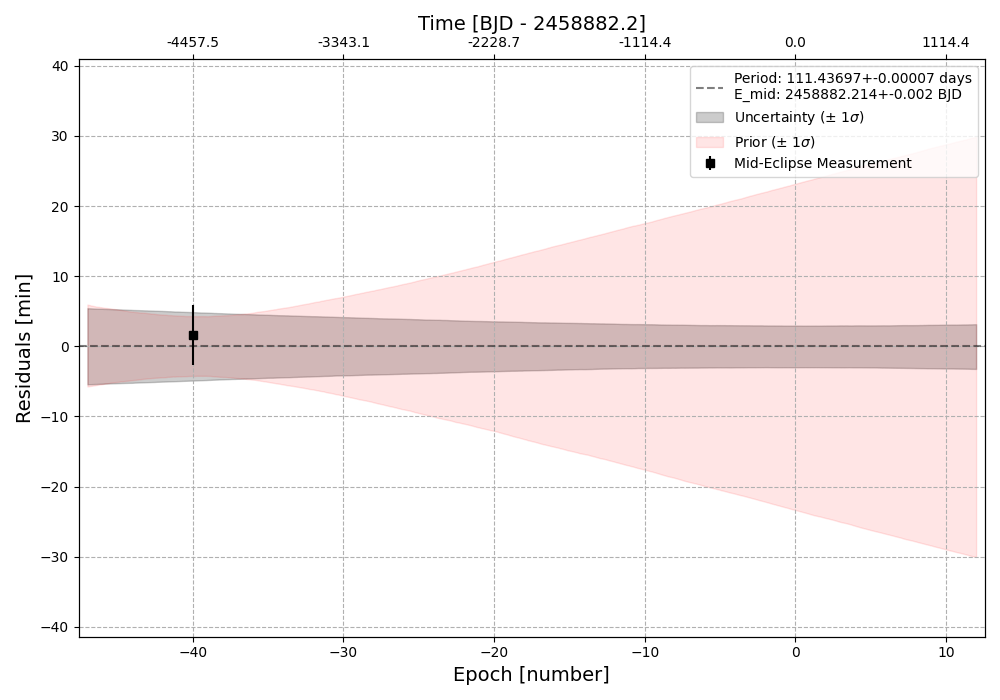}
\caption{ left) A comparison of residuals between the measured mid-transit times and a calculated linear ephemeris (reported in the plot legend). The grey shaded region indicates the uncertainty in the ephemeris extending to $\pm$1 $\sigma$ using our best estimates in Table \ref{tab:final}. The pink shaded region indicates an uncertainty based on the prior listed in Table \ref{tab:nominal}. Some mid-transit measurements are not used in the joint analysis because they were measured from partial transits. right) An ephemeris estimate for mid-eclipse times. The pink shaded region shows the uncertainty in a linear solution if we use the Spitzer measurement \citep{Laughlin2009} as $E_{mid}$ along with the period from \citealt{Bonomo2017}. The grey shaded region indicates an uncertainty based on the orbital information listed in Table \ref{tab:final}.
\label{fig:period}}
\end{figure*}

\section{Results and Conclusions \label{sec:params}}

As part of an effort to refine the orbital ephemeris for HD 806060 b, we have obtained  new radial velocity and transit measurements for HD~80606~b. The transit measurements were obtained with TESS in 2020 and a ground-based campaign in 2021; together, the new data, coupled with archival RV and transit observations provide a valuable constraint on the time of conjunction. We are able to refine the estimate on the orbital period of HD~80606~b by taking advantage of the 10-year baseline between the archival and  the new observations. Using only the data from 2009-2010, the uncertainty on the orbital period was $\sigma$(P) $=4\times 10^{-4}$; combining the old data with the new observations, the new value of the period 111.436971~days has an improved uncertainty of $\sigma$(P)= $7.4\times 10^{-5}$~days (Figure~\ref{fig:period}). The period estimate is improved by factor $\sim$5 compared to \cite{Bonomo2017} along with significant improvements for the system parameters as summarized in Table~\ref{tab:final}. The immediate benefit of these new observations is to greatly reduce the uncertainty in the timing of future events (transits or eclipses; e.g., \citealt{Zellem2020}). 

In the case of an eclipse in November 2022, e.g., in mid Cycle 1 for JWST, the uncertainty resulting from propagating the ephemeris in Table~\ref{tab:nominal} is $\sim$24 minutes, whereas with the new linear ephemeris  the uncertainty is $\sim$5 minutes (See Figure~\ref{fig:period}). The linear ephemeris uses the eclipse mid-point from \citep{Laughlin2009} and our new period estimate. We also provide a more conservative error estimate based on the orbit solution which yields an uncertainty $\sim$30 minutes. The orbit solution has a larger uncertainty than the linear propagation due to the uncertainty in $e$ and $\omega$ on the estimated eclipse time. For example, the mid-eclipse time predicted from the prior is 2458882.207 $\pm$ 0.10 and from our posterior we get 2458882.214 $\pm$ 0.021 which leads to a difference in uncertainty of $\sim$2 hours. The errors are significantly larger on predicting mid-eclipse because of a degeneracy between $e$ and $\omega$ and it is exacerbated with larger orbital periods. Removing the degeneracy may be possible by simultaneously fitting a transit and eclipse. The uncertainties reported in Figure $\ref{fig:period}$ are smaller than the ones estimated above because they use a linear propagation of the average orbit solution. It is also important to note that the uncertainty on inclination in the prior does not always yield a transiting planet when conducting a Monte Carlo simulation. Simultaneously fitting a TESS light curve with RV data allowed for a strong constraint on the inclination that helped measure the transit duration to within $\sim$ 7 minutes compared to the full event that is almost 12 hours.

For the analysis of the JWST phase curve it is important to know the offset between the eclipse, which will be well determined by the JWST observations, and time of periapsis, which will not be directly measured. The timing of eclipse relative to periapsis depends on three key variables: orbital period $P$, eccentricity $e$, and argument of periapsis $\omega$ in Eqn~(\ref{deltaT}) \citep{Huber2017, Alonso2018}: 
\begin{equation}
    T_{ecl}-T_{peri}=  \frac{P}{2 \pi \sqrt{
     1 - e^2}} \int_{0}^{-\frac{\pi}{2}
     -\omega} \left( \frac{(1 - e^2)}{1 + e cos(x)} \right)^2 \,dx \label{deltaT}
\end{equation}

A Monte Carlo simulation for the parameters with their associated uncertainties (Table~\ref{tab:final}) yields an offset in time between the eclipse and periapsis of $\Delta T =-3.104\pm0.011$ hr, i.e with the eclipse occurring before periapsis. This is to be compared with -3.069$\pm$0.049 hr  derived using the \citet{Bonomo2017} parameters in Table~\ref{tab:nominal}, a difference of $\sim$2 minutes. Table~\ref{tab:ephemeris} takes the times of periapsis, eclipse and conjunction from our solution (Table~\ref{tab:final}) and propagates these forward in time from 2020 to 2031. The uncertainties include a constant term from the initial Monte Carlo estimates plus the growth in uncertainty occurring $N$ periods after the reference time.

Finally, we note that the increased precision of the ephemeris, when combined with new JWST observations, may allow an exploration of non-Keplerian effects such as tidal dissipation \citep{Fabrycky2010} or General Relativistic effects similar to those seen in the precession of the periapsis in orbit of Mercury in our solar system, but greatly enhanced by the high eccentricity of HD~80606~b. \citet{Blanchet2019} calculate that offsets between transit and eclipse midpoints should grow as the number of orbits increases. While the precision and temporal baseline of the 2009--2010 measurements is inadequate to measure the predicted effects of 3--4 minutes, the high precision expected from JWST's great sensitivity make such measurements possible over the next few years. Additionally, our measurements reported in this paper will be archived on ExoFOP enabling future studies to search for long-term perturbations that may affect the ephemeris estimates.

\begin{deluxetable*}{llccc}
\centering
\tablecaption{System Parameters for HD~80606\label{tab:final}}
\tablehead{
\colhead{Parameter} & \colhead{Explanation} & \colhead{Our Study} & \cite{Rosenthal2021} & \cite{Bonomo2017} }
\startdata
M$_*$ [M$_\odot$]   & Stellar Mass   & 1.05  & 1.047$\pm$0.047 & 1.018$\pm$0.035 \\
R$_*$ [R$_\odot$]   & Stellar Radius & 1.050 $\pm$ 0.01$^a$ &  1.066$\pm$0.024 & 1.037$\pm$0.032 \\
T$_*$ [K]           & Stellar Temperature & 5565 & 5565 $\pm$ 92 & 5574$\pm$72 \\
Fe/H                & Stellar Metallicity & 0.35 & 0.348$\pm$0.057 & 0.340$\pm$0.050 \\
$(R_{p}/R_*)_{contaminated}$ & Planet-Star Radius Ratio & 0.07268 $\pm$ 0.00085 \\
$(R_p/R_*)^2_{contaminated}$ & Radius Ratio Squared & 0.00528 $\pm$ 0.00012 \\ 
$(R_p/R_*)^2_{corrected}$ & Radius Ratio Squared  & 0.01019 $\pm$ 0.00023$^b$ &  & 0.00991$\pm$0.00076  \\
$R_p$ [R$_{Jupiter}$] & Planet Radius  & 1.032$\pm$0.015 &  & 1.003$\pm$0.023  \\
$M_{p}$ [M$_{Jupiter}$] & Planet Mass & 4.1641 $\pm$ 0.0047 & 4.16 $\pm$0.13$^c$ & 4.1$\pm$0.1 \\
K [m/s]             & RV Semi-Amplitude & 469.22 $\pm$ 0.61 & 465.5$\pm$2.8 & 474.9$\pm$2.6 \\
Period [day]        & Orbital period & 111.436765 $\pm$ 0.000074 & 111.43639$\pm$0.00032 & 111.4367$\pm$0.0004  \\
E$_{mid}$ [BJD]     & Eclipse Midpoint & 2458882.214 $\pm$ 0.0021$^d$ &  & \\
E$_{14}$ [day]      & Eclipse Duration &  0.07169$\pm$0.00073 &  &  \\
T$_{peri}$ [BJD]    & Epoch of periastron & 2458882.344 $\pm$ 0.0021 &  &  \\
T$_{mid}$ [BJD]     & Transit Midpoint & 2458888.07466 $\pm$ 0.00204 & 2455099.39$\pm$0.13 & 2455210.6428$\pm$0.001 \\
T$_{14}$ [day]      & Transit Duration & 0.4990 $\pm$ 0.0048 &  & \\
$i$ [deg]           & Inclination & 89.24 $\pm$ 0.01 &  & 89.23$\pm$0.3 \\
a/R$_*$             & Scaled Semi-major axis & 94.452 $\pm$ 0.014 & 92.8$\pm$2.5 & 94.6$\pm$3.1 \\
a [au]              & Semi-major axis & 0.4603$\pm$0.0021 & 0.4602$\pm$0.0071 & 0.4565$\pm$0.0053 \\
$e$                 & Eccentricity &  0.93183 $\pm$ 0.00014  & 0.93043$\pm$0.00068 & 0.93226$\pm$0.00064 \\
$\omega$ [deg]      & Arg. of periastron & -58.887 $\pm$ 0.043  & -58.95$\pm$0.25 & -58.97$\pm$0.2 \\
\enddata
\tablecomments{The values in parentheses are calculated using the respective column's orbit solution and a Monte Carlo simulation with 10,000 forward model evaluations. $^a$Gaussian Prior; $^b$Corrected for stellar contamination using brightness values for HD~80606: V-mag=9.00 and HD80607: V-mag=9.07; $^c$ M$_{p}$sin($i$); $^d$Uncertainty estimated with fixed $\omega$;}
\end{deluxetable*}

\begin{deluxetable*}{lllll}
\tabletypesize{\scriptsize}
\tablecaption{Predicted Transit, Eclipse and Periapsis Times \label{tab:ephemeris}}
\tablehead{
\colhead{Period} & \colhead{Periapsis Date}& \colhead{T$_{Peri}$ (BJD$_{TBD}$)}& \colhead{E$_{mid}$ (BJD$_{TBD}$)}
& \colhead{T$_{mid}$ (BJD$_{TBD}$)}}
\startdata
0 & 2020-02-02 20:15:10 & 2458882.344 $\pm$ 0.0021 & 2458882.214 $\pm$ 0.0021 & 2458888.0746 $\pm$ 0.0020 \\
1 & 2020-05-24 06:44:36 & 2458993.781 $\pm$ 0.0021 & 2458993.651 $\pm$ 0.0021 & 2458999.5116 $\pm$ 0.0020 \\
2 & 2020-09-12 17:14:03 & 2459105.218 $\pm$ 0.0021 & 2459105.089 $\pm$ 0.0021 & 2459110.9487 $\pm$ 0.0020 \\
3 & 2021-01-02 03:45:18 & 2459216.656 $\pm$ 0.0021 & 2459216.527 $\pm$ 0.0021 & 2459222.3855 $\pm$ 0.0021 \\
4 & 2021-04-23 14:12:08 & 2459328.092 $\pm$ 0.0021 & 2459327.962 $\pm$ 0.0021 & 2459333.8225 $\pm$ 0.0021 \\
5 & 2021-08-13 00:42:14 & 2459439.529 $\pm$ 0.0022 & 2459439.400 $\pm$ 0.0022 & 2459445.2595 $\pm$ 0.0022 \\
6 & 2021-12-02 11:10:00 & 2459550.965 $\pm$ 0.0022 & 2459550.836 $\pm$ 0.0022 & 2459556.6963 $\pm$ 0.0021 \\
7 & 2022-03-23 21:40:27 & 2459662.403 $\pm$ 0.0021 & 2459662.274 $\pm$ 0.0022 & 2459668.1333 $\pm$ 0.0021 \\
8 & 2022-07-13 08:09:58 & 2459773.840 $\pm$ 0.0022 & 2459773.711 $\pm$ 0.0021 & 2459779.5704 $\pm$ 0.0021 \\
9 & 2022-11-01 18:39:52 & 2459885.278 $\pm$ 0.0023 & 2459885.148 $\pm$ 0.0022 & 2459891.0073 $\pm$ 0.0022 \\
10 & 2023-02-21 05:08:03 & 2459996.714 $\pm$ 0.0022 & 2459996.584 $\pm$ 0.0023 & 2460002.4443 $\pm$ 0.0022 \\
11 & 2023-06-12 15:37:31 & 2460108.151 $\pm$ 0.0023 & 2460108.021 $\pm$ 0.0021 & 2460113.8814 $\pm$ 0.0022 \\
12 & 2023-10-02 02:07:17 & 2460219.588 $\pm$ 0.0023 & 2460219.459 $\pm$ 0.0022 & 2460225.3183 $\pm$ 0.0022 \\
13 & 2024-01-21 12:36:17 & 2460331.025 $\pm$ 0.0022 & 2460330.896 $\pm$ 0.0022 & 2460336.7554 $\pm$ 0.0023 \\
14 & 2024-05-11 23:06:49 & 2460442.463 $\pm$ 0.0024 & 2460442.334 $\pm$ 0.0023 & 2460448.1923 $\pm$ 0.0023 \\
15 & 2024-08-31 09:35:05 & 2460553.899 $\pm$ 0.0023 & 2460553.770 $\pm$ 0.0023 & 2460559.6291 $\pm$ 0.0023 \\
16 & 2024-12-20 20:01:54 & 2460665.335 $\pm$ 0.0023 & 2460665.205 $\pm$ 0.0024 & 2460671.0663 $\pm$ 0.0023 \\
17 & 2025-04-11 06:33:17 & 2460776.773 $\pm$ 0.0024 & 2460776.644 $\pm$ 0.0024 & 2460782.5030 $\pm$ 0.0023 \\
18 & 2025-07-31 17:03:20 & 2460888.211 $\pm$ 0.0024 & 2460888.081 $\pm$ 0.0025 & 2460893.9400 $\pm$ 0.0025 \\
19 & 2025-11-20 03:30:27 & 2460999.646 $\pm$ 0.0025 & 2460999.517 $\pm$ 0.0024 & 2461005.3771 $\pm$ 0.0024 \\
20 & 2026-03-11 14:00:39 & 2461111.084 $\pm$ 0.0024 & 2461110.954 $\pm$ 0.0024 & 2461116.8140 $\pm$ 0.0025 \\
21 & 2026-07-01 00:29:17 & 2461222.520 $\pm$ 0.0024 & 2461222.391 $\pm$ 0.0025 & 2461228.2509 $\pm$ 0.0025 \\
22 & 2026-10-20 10:59:11 & 2461333.958 $\pm$ 0.0026 & 2461333.828 $\pm$ 0.0025 & 2461339.6880 $\pm$ 0.0025 \\
23 & 2027-02-08 21:26:36 & 2461445.393 $\pm$ 0.0025 & 2461445.264 $\pm$ 0.0026 & 2461451.1249 $\pm$ 0.0026 \\
24 & 2027-05-31 07:56:26 & 2461556.831 $\pm$ 0.0025 & 2461556.701 $\pm$ 0.0026 & 2461562.5616 $\pm$ 0.0027 \\
25 & 2027-09-19 18:27:19 & 2461668.269 $\pm$ 0.0026 & 2461668.140 $\pm$ 0.0026 & 2461673.9988 $\pm$ 0.0026 \\
26 & 2028-01-09 04:54:59 & 2461779.705 $\pm$ 0.0027 & 2461779.575 $\pm$ 0.0027 & 2461785.4358 $\pm$ 0.0027 \\
27 & 2028-04-29 15:27:05 & 2461891.144 $\pm$ 0.0028 & 2461891.014 $\pm$ 0.0027 & 2461896.8728 $\pm$ 0.0028 \\
28 & 2028-08-19 01:54:49 & 2462002.580 $\pm$ 0.0028 & 2462002.450 $\pm$ 0.0028 & 2462008.3097 $\pm$ 0.0029 \\
29 & 2028-12-08 12:23:09 & 2462114.016 $\pm$ 0.0029 & 2462113.887 $\pm$ 0.0029 & 2462119.7467 $\pm$ 0.0030 \\
30 & 2029-03-29 22:52:37 & 2462225.453 $\pm$ 0.0030 & 2462225.324 $\pm$ 0.0029 & 2462231.1838 $\pm$ 0.0031 \\
31 & 2029-07-19 09:22:28 & 2462336.891 $\pm$ 0.0030 & 2462336.761 $\pm$ 0.0031 & 2462342.6207 $\pm$ 0.0031 \\
32 & 2029-11-07 19:51:46 & 2462448.328 $\pm$ 0.0032 & 2462448.198 $\pm$ 0.0030 & 2462454.0576 $\pm$ 0.0030 \\
33 & 2030-02-27 06:22:46 & 2462559.766 $\pm$ 0.0031 & 2462559.636 $\pm$ 0.0032 & 2462565.4946 $\pm$ 0.0031 \\
34 & 2030-06-18 16:50:58 & 2462671.202 $\pm$ 0.0032 & 2462671.072 $\pm$ 0.0032 & 2462676.9315 $\pm$ 0.0031 \\
35 & 2030-10-08 03:19:59 & 2462782.639 $\pm$ 0.0033 & 2462782.509 $\pm$ 0.0033 & 2462788.3685 $\pm$ 0.0032 \\
\enddata
\end{deluxetable*}

\section{Acknowledgements}

Some of the research described in this publication was carried out in part at the Jet Propulsion Laboratory, California Institute of Technology, under a contract with the National Aeronautics and Space Administration. This research has made use of the NASA Exoplanet Archive and ExoFOP, which is operated by the California Institute of Technology, under contract with the National Aeronautics and Space Administration under the Exoplanet Exploration Program.

This publication makes use of data products from Exoplanet Watch, a citizen science project managed by NASA's Jet Propulsion Laboratory on behalf of NASA's Universe of Learning. This work is supported by NASA under award number NNX16AC65A to the Space Telescope Science Institute, in partnership with Caltech/IPAC, Center for Astrophysics|Harvard $\&$ Smithsonian, and NASA Jet Propulsion Laboratory.

We acknowledge with thanks the use of the AAVSO Exoplanet Database contributed by observers worldwide and used in this research.

This work makes use of observations from the Las Cumbres Observatory global telescope network. The authors thank Dr. Lisa Storie-Lombardi for the grant of Director's Discretionary Time with the Los Cumbres Observatory (LCO) which was critical to the execution of this program. Dr. Rachel Street  helped to identify the appropriate telescopes and observing modes for LCO.

Some of the data presented herein were obtained at the W. M. Keck Observatory, which is operated as a scientific partnership among the California Institute of Technology, the University of California and the National Aeronautics and Space Administration. The Observatory was made possible by the generous financial support of the W. M. Keck Foundation.  The authors wish to recognize and acknowledge the very significant cultural role and reverence that the summit of Maunakea has always had within the indigenous Hawaiian community.  We are most fortunate to have the opportunity to conduct observations from this mountain.

Some of the scientific data presented herein were obtained using the eVscope Network, which is managed jointly by Unistellar and the SETI Institute. The Unistellar Network and work by T.M.E. and A.A. are supported by grants from the Gordon and Betty Moore Foundation. The authors wish to thank Prof. S. Kulkarni for an introduction to members of the GROWTH consortium,

The results reported herein benefited from collaborations and/or information exchange within NASA's Nexus for Exoplanet System Science (NExSS) research coordination network sponsored by NASA's Science Mission Directorate.

K.W. acknowledges support from NASA through the NASA Hubble Fellowship grant HST-HF2-51472.001-A awarded by the Space Telescope Science Institute, which is operated by the Association of Universities for Research in Astronomy, Incorporated, under NASA contract NAS5-26555.

This research has made use of the NASA Exoplanet Archive, which is operated by the California Institute of Technology, under contract with the National Aeronautics and Space Administration under the Exoplanet Exploration Program.

The ExoClock project has received funding from the UKSA and STFC grants: ST/W00254X/1 and ST/W006960/1.

This work made use of data from the GROWTH-India Telescope (GIT) set up by the Indian Institute of Astrophysics (IIA) and the Indian Institute of Technology Bombay (IITB). It is located at the Indian Astronomical Observatory (Hanle), operated by IIA. We acknowledge funding by the IITB alumni batch of 1994, which partially supports operations of the telescope. Telescope technical details are available at \url{https://sites.google.com/view/growthindia/}.

This work uses funding from the Ministry of Science and Technology (Taiwan) under the contract 109-2112-M-008-014-MY3 and we are thankful for their support. The queue observations were done using the 0.4m SLT telescope located at the Lulin Observatory, with assistance from observatory staff C.-S. Lin, H.-Y. Hsiao, and W.-J. Hou.

\facility{Keck:I (HIRES), Lick:APF, LCO, TESS, Spitzer Space Telescope, Keck Observatory Archive (KOA)}


\appendix

The appendix contains figures regarding the best-fit solution for the ground-based transit light curve along with posteriors from the analysis. 

\begin{turnpage}
\begin{figure*}[h!]
\centering
\includegraphics[width=1.2\textwidth]{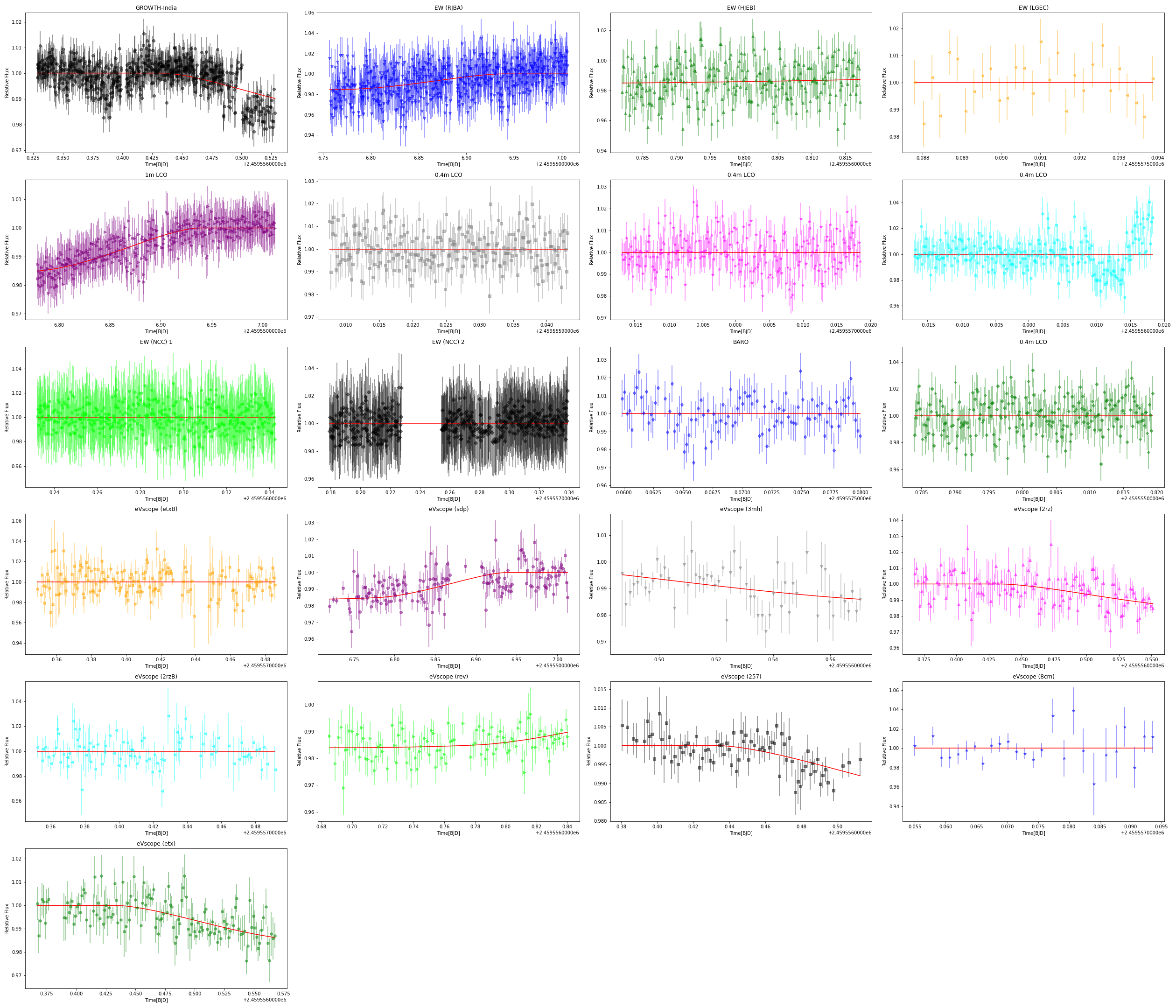} \\
\caption{A mosaic of observations for the transit of HD 80606 b on the night of 7--8 December 2021. A global light curve model is over-plotted in red. All of these observations are stitched together into a single time series shown in Figure \ref{fig:BestFit}.
\label{fig:mosaic}}
\end{figure*}
\end{turnpage}

\begin{figure*}[b!]%
\centering
\includegraphics[width=0.99\textwidth]{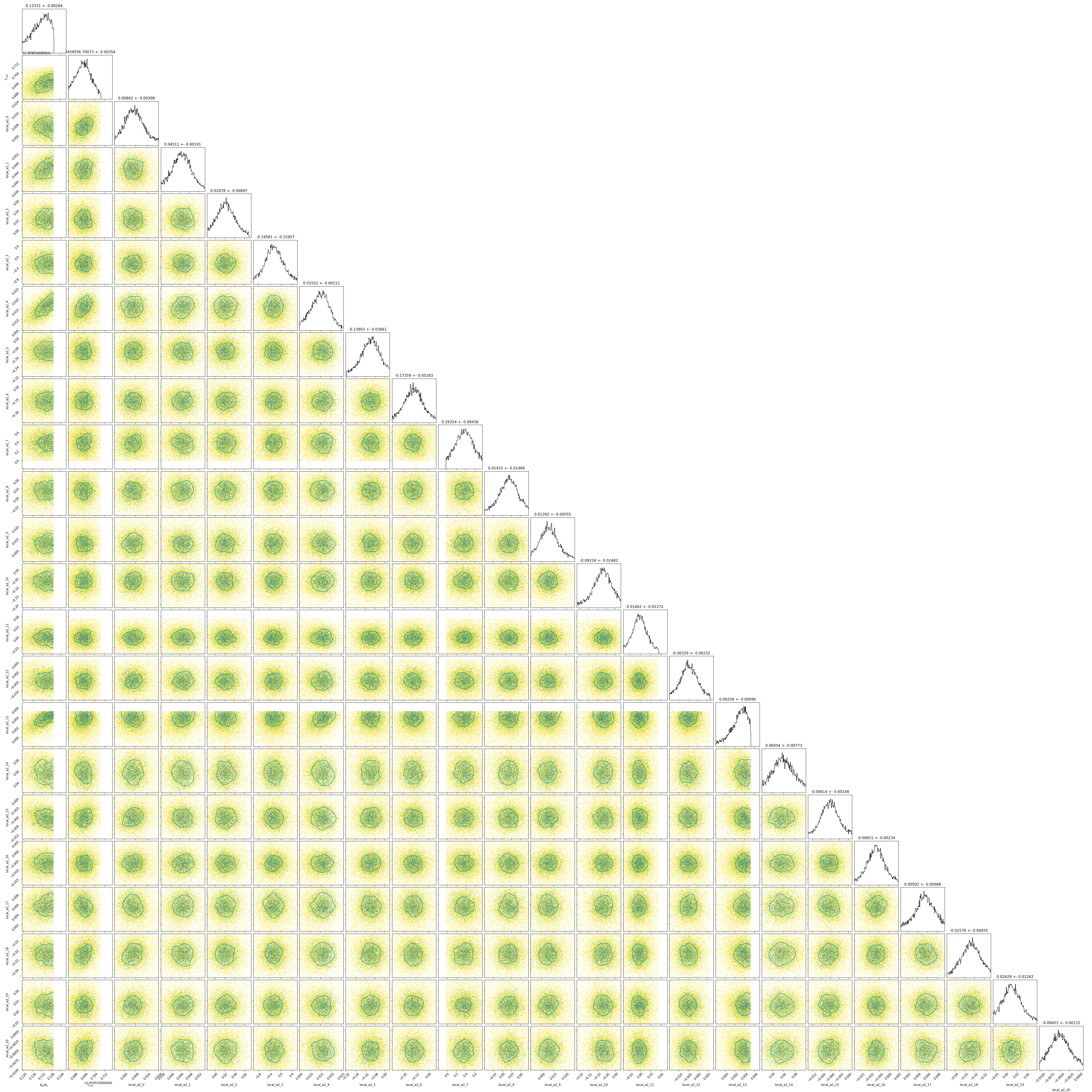} \\
\caption{Posteriors for the global light curve solution on 7--8 December 2021. All of the observations shown in Figure \ref{fig:BestFit} are fit simultaneously in order to constrain the time of mid-transit and transit depth from a global light curve model. Each observation was acquired at a different airmass and requires individual treatment in order to detrend properly and those coefficients make up a majority of this distribution. The data points in each correlation plot are color coded to the likelihood with darker colors representing higher likelihoods. The contour represents roughly the 1 $\sigma$ boundary using the uncertainty reported in each plot's title. Each values in the prior starts as a uniform distribution.
\label{fig:posterior}}
\end{figure*}

\begin{figure*}[b!]%
\centering
\includegraphics[width=0.9\textwidth]{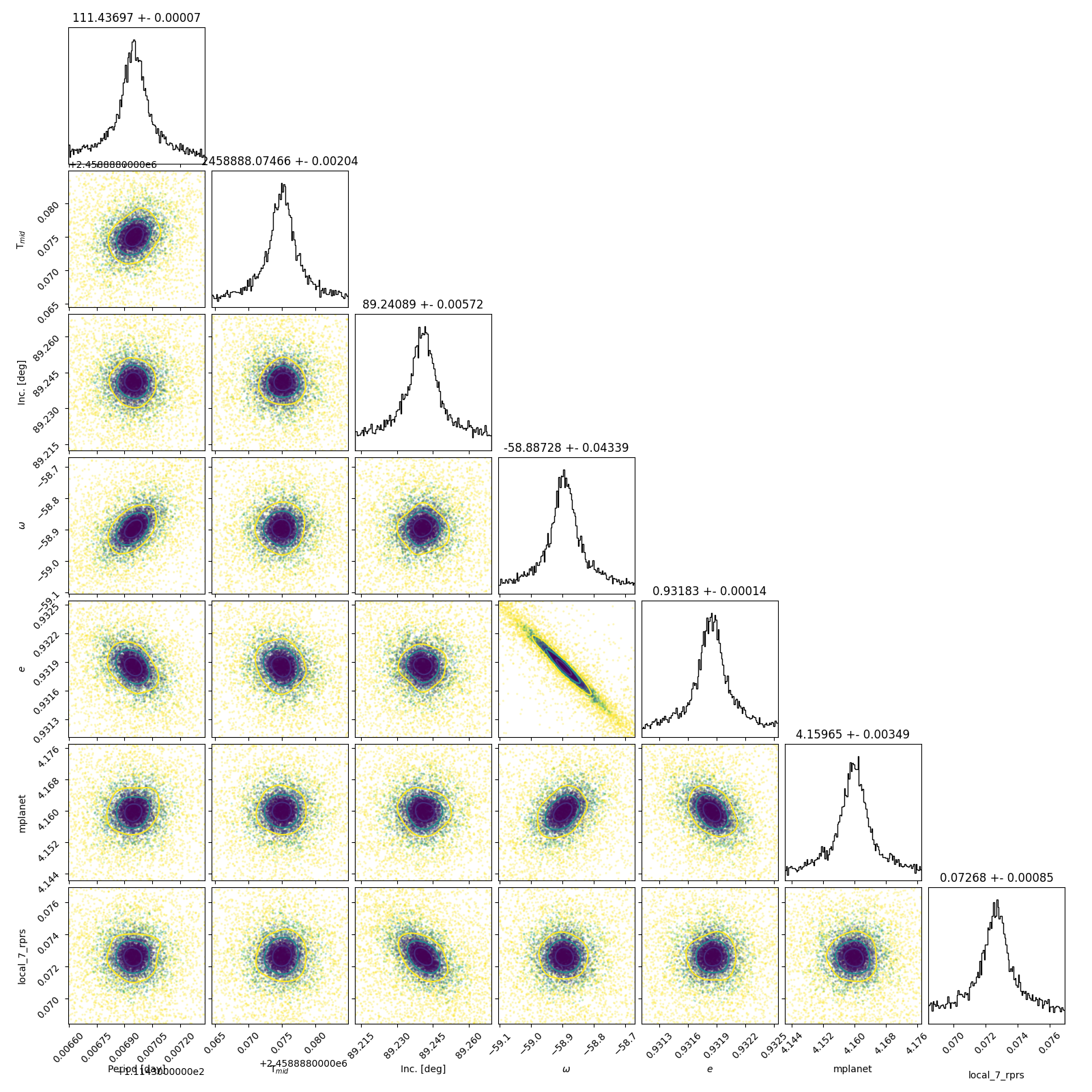}
\caption{Posteriors for the joint fit between the radial velocity measurements, a transit light curve from TESS and historical measurements for mid-transit and mid-eclipse. Measurements from Table~$\ref{tab:oldmid}$ are included in an ephemeris estimate during the fitting process and added into the likelihood function. The best-fit radial velocity model can be found in Figure \ref{fig:joint_rv}, the best-fit transit model is shown in Figure \ref{fig:joint_transit} and the final ephemeris is shown in Figure \ref{fig:period}. The data points in each correlation plot are color coded to the likelihood with darker colors representing higher likelihoods. The contours represent the N$\sigma$ boundary using the uncertainty reported in each column's title.
\label{fig:joint_posteriors}}
\end{figure*}

\end{CJK*}
\end{document}